\documentclass[11pt]{article}
\makeatletter
\@addtoreset{equation}{section}

\makeatother

\begin{document}

\begin{titlepage}

\renewcommand{\thefootnote}{\fnsymbol{footnote}}

\hfill\parbox{4cm}{hep-th/0411006  \\ KIAS-P04042}
\vspace{15mm}
\baselineskip 9mm
\begin{center}
{\LARGE \bf Type 0A matrix model of black hole, integrability and holography}

\end{center}

\baselineskip 6mm
\vspace{10mm}
\begin{center}
Jaemo Park$^a$\footnote{\tt jaemo@physics.postech.ac.kr} 
and Takao Suyama$^b$\footnote{\tt suyama@kias.re.kr} 
\\[5mm] 
{\sl $^a$ Department of Physics, POSTECH, Pohang 790-784,
Korea \\ 
$^b$ School of Physics, Korea Institute for Advanced Study,
Seoul 130-722, Korea}
\end{center}

\thispagestyle{empty}

\vfill
\begin{center}
{\bf Abstract}
\end{center}
\noindent
We investigate a deformed matrix model of type 0A theory related to supersymmetric 
Witten's black hole in two-dimensions, generalization of
bosonic model suggested by Kazakov et. al. We find a free field realization 
of the partition function of the matrix model, which includes Ramond-Ramond 
perturbations in the type 0A theory. In a simple case, the partition function 
is factorized into two determinants, which are given by $\tau$ function of an 
integrable system. We work out the genus expansion of the partition function.
Holographic relation with the supersymmetric Witten's black hole is checked 
by Wilson line computation. Corresponding partition function of the matrix model 
exhibits a singular behavior, which is interpreted as the point of enhanced 
${\cal N}=2$ worldsheet supersymmetry. Interesting relation of the deformed matrix model 
and topological string on a $Z_2$ orbifold of conifold is found.

\vspace{20mm}
\end{titlepage}

\vspace{1cm}

\section{Introduction}

Kazakov, Kutasov and Kostov \cite{KKK}  propose a matrix model dual to the 
Sine-Liouville theory, which in turn is conjectured to be 
equivalent to the 2-D string theory admitting black hole 
background described by a coset model of $SL(2,R)/U(1)$ type. 
This is known as the Witten's 2-D black hole\cite{Witten}. 
Sine-Liouville theory is the deformation of the usual Liouville 
theory by winding modes or the condensation of vortices.
In the matrix model side, winding mode perturbations are described 
as the insertion of exponentiated Wilson line operator.
In \cite{SY}, it is shown that the duality relation between the 2-D string theory 
on the Witten's black hole and the deformed matrix model could 
be understood in terms of {\it gauge theory/gravity theory correspondence}.
In \cite{SY}, the proposed matrix model of Kazakov, Kostov and Kutasov 
\cite{KKK}
 was shown to be a gauge theory.
This reparaphrased holography was checked by computing the Wilson line
expectation value and comparing it with the bulk computation. This check 
of the holographic relation was made in the context of the bosonic 
string.

One motivation of this paper is to understand the holographic 
relation in the context of the type 0A string. Obviously one can 
consider the Witten's black hole with ${\cal N}=1$ worldsheet supersymmetry
in the type 0A setup. 
In addition,  if we consider the 2-D supergravity 
action describing low energy theory of the type 0A string theory,
we can have a classical 2-D extremal black hole solution\cite{Berkovits}.
This is argued to be dual to the type 0A matrix model with $\mu=0$ and 
nonzero $f$, where $\mu$ denotes the usual Fermi sea level and $f$ 
is related to the Ramond-Ramond(RR) flux in the type 0A 
string theory\cite{Takayanagi}. Also it is conjectured that non-extremal black holes 
are described by the Wilson line deformation of the type 0A matrix 
model\cite{Takayanagi}. Thus it might be interesting 
to work out
the corresponding matrix models or gauge theories holographically 
dual to these black hole solutions of the type 0A string theory. 
In \cite{KKK}, the partition function of the underlying matrix model 
describing $SL(2,R)/U(1)$ black hole background is given by the $\tau$
function of an integrable system. We might guess that similar 
integrable structures underlie on the type 0A matrix model dual to
2-D black hole backgrounds. We consider the type 0A matrix model with 
exponentiated Wilson line operators which correspond to two kinds of 
winding mode perturbations in the type 0A Liouville theory.
Indeed we find a free field realization of the partition function
of the type 0A matrix model and find that turning on RR flux in the string 
theory corresponds to the nonzero relative momentum in the free 
field representation. However the analogue of the Hirota bilinear 
equation, which would give the needed differential equation governing 
the partition function, is more complicated than the bosonic case. 

The situation is greatly simplified if we consider turning on no RR flux 
and turning on the same perturbations for both winding modes. This is again 
related to the Witten's black hole solution with ${\cal N}=1$ worldsheet supersymmetry.
In this case the partition function is decomposed into two determinants, 
which are described by the $\tau$ function again.
The resulting partition function shows singular behavior at two 
points of the parameter spaces, which indicates the vanishing of the 
mass gap or the emergence of the new degrees of freedom in the infrared.
One of the singularity is well known. The analogous situation in the bosonic
case is related to the conifold singularity of a Calabi-Yau space. 
Using the relation between 
$c=1$ matrix model at the self-dual radius and topological string 
on the conifold\cite{conifold}, this corresponds to vanishing of the 3-cycle 
of the conifold and black hole state of the string theory on the conifold 
becomes massless\cite{Strominger}. From the conformal field theory side, this
means 
that the mass gap of the conformal field theory(CFT) vanishes\cite{OV}. 
The other singularity is not known
so far.
The singularity structure is important since the holographic relation 
holds for this point of the parameter space, as indicated 
by the Wilson line calculations. We suggest that this singularity is 
related to the enhancement of the worldsheet supersymmetry of ${\cal N}=1 \,\,
SL(2,R)/U(1)$ theory to ${\cal N}=2$ worldsheet supersymmetry. 

Another interesting point is that the compactified radius of the type 0A
matrix model indicates that the theory is equivalent to some topological 
string, which is identified with the IIB topological string on a $Z_2$ 
orbifold of the conifold. It is well known that $c=1$ matrix model 
at the self-dual radius is equivalent to the IIB topological string 
theory on the conifold\cite{conifold}. Here we see that type 0A version of 
this relation
is going to be useful in understanding the holographic relation.
According to \cite{hierarchy}, the underlying integrable structure of $c=1$ 
matrix model can be encoded into large symmetries of the conifold.
This approach would be useful in elucidating the integrable structures 
of the type 0A matrix models of our interest. Related issues are also discussed
in\cite{Olsson} 

The content of the paper is as follows.
In section 2, we briefly review the various equivalences between two-dimensional 
CFT describing Witten's black hole, Sine-Liouville theory and the corresponding 
matrix model for bosonic case and introduces some preliminaries needed for 
type 0A generalizations.
In section 3, we introduce a deformed matrix model related to ${\cal N}=1$ 
2-D black hole. 
In section 4, we work out the determinant representation and the 
factorization of the partition function in a simple case. 
In section 5, we explicitly work out the free energy of the type 0A 
matrix model. In section 6,we carry out Wilson line computation and 
in section 7, we comment on the relation to topological string.
In appendices, determinant representation and free field representation
of the partition function are worked out.

\vspace{1cm}

\section{Review of bosonic case and preliminaries for type 0A case}

\vspace{5mm}

Witten's two-dimensional black hole is described by $SL(2, R)/U(1)$ coset 
conformal field theory with level $k=\frac{9}{4}$\cite{Witten}. Its two-dimensional 
geometry is given by 
\begin{eqnarray}
ds^2 &=&\alpha ' k (dr^2+\tanh^2 r d\theta^2)  \label{bh}\\
\Phi&=&\Phi_0-\frac{1}{2}\log (\cosh 2r +1)      \nonumber
\end{eqnarray}
where $\Phi$ is the usual dilaton field.
This is derived by considering the bare Lagrangian of gauged WZW 
model of $SL(2,R)/U(1)$. For bosonic case there are additional $\alpha '$
corrections\cite{DVV}\cite{Tseytlin}. According to the conjectures made by Fateev,Zamolodchikov and 
Zamolodchikov\cite{FZZ}, this coset CFT 
is equivalent to the Sine-Liouville theory 
\begin{equation}
S=\frac{1}{4\pi}\int d^2\sigma [ (\partial x)^2+(\partial\phi)^2-Q\hat{R}\phi
+\lambda e^{b\phi}\cos R(x_L-x_R)]
\end{equation}
with $R=\frac{3}{2}, Q=2$ and $b=-\frac{1}{2}$.
In the paper \cite{KKK}, the matrix model equivalent to the Sine-Liouville theory
is constructed. We consider the $c=1$ matrix quantum mechanics compactified 
on a circle with radius $R$ with a twisted boundary condition, whose partition 
function is given by
\begin{equation}
Z_N(\Omega)=\int_{M(2\pi R)=\Omega^{\dagger}M(0) \Omega} {\cal D}M(x)
\exp(-Tr \int_0^{2\pi R} dx [ \frac{1}{2} (\partial_x M)^2+V(M)]),
\end{equation}
where $M(x)$ is an $N\times N$ matrix-valued field on the circle and $\Omega$
is a unitary matrix. The form of $V(M)$ is 
\begin{equation}
V(M)=\frac{1}{2}M^2-\frac{g}{3\sqrt{N}}M^3
\end{equation}
but the precise form of the potential is not important in the double 
scaling limit. The proposed matrix model is defined by the partition function
\begin{equation}
Z_{N}(\lambda)=\int {\cal D}\Omega 
e^{\lambda Tr(\Omega+\Omega^{\dagger})}Z_N(\Omega).  \label{matrix}
\end{equation}
In \cite{KKK}, it is shown that the matrix model (\ref{matrix}) is 
equivalent to a 
modified version of the Sine-Liouville theory
\begin{equation}
S=\frac{1}{4\pi}\int d^2\sigma [ (\partial x)^2+(\partial\phi)^2
-2\hat{R}\phi+\mu e^{-2\phi}
+\lambda e^{(R-2)\phi}cos R(x_L-x_R)].
\end{equation}
In a suitable limit where the term $\mu e^{-2\phi}$ is negligible, the modified 
theory reduces to the Sine-Liouville theory, which is equivalent to the black 
hole 
CFT.

We consider the generalization to the theories with ${\cal N}=1$ worldsheet 
supersymmetry.
Type 0A Sine-Liouville theory contains two worldsheet superfield $X$ 
with central charge 
$c=\frac{3}{2}$ and $\phi$ with $c=\frac{3}{2}+6Q^2$. The total central charge
$c=15$ determines $Q=\sqrt{2}$. On the other hand, ${\cal N}=1 \, SL(2, R)/U(1)$ CFT 
with 
level $k$ has the total central charge $c=\frac{3(k+2)}{k}$. From $c=15$ we 
obtain $k=\frac{1}{Q^2}=\frac{1}{2}$. The two-dimensional geometry (\ref{bh}) 
is not renormalized with ${\cal N}=1$ worldsheet supersymmetry\cite{9201039}. 
Hence the compactification 
radius $R=\sqrt{\alpha ' k}$
and we obtain $R=\frac{1}{2}$ with the convention $\alpha'=\frac{1}{2}$ 
and this is the same compactification radius of the 
time direction $X$ 
for the matrix model if the holographic relation holds. 
When we compactify $X$ with the radius $R$, we can have the superaffine theory 
as well as the usual type 0A theory. These two theories differ by the 
GSO projections\cite{newhat}.
While the superaffine theory correlates the sum over spin structures with the 
momentum 
and winding of $X$, the type 0A theory does not. For the supercoset model, it is 
natural to consider the GSO projections which do not correlate the fermionic 
projections 
with the momentum and winding modes, which we call diagonal GSO projections. Thus 
we look for the holographic relation
between the type 0A theory and ${\cal N}=1$ supercoset with the diagonal GSO projections.
For the type 0A case, matrix model should be suitably generalized and this will be 
explained in the next section.

\vspace{1cm}
\section{Matrix model of type 0A string}

\vspace{5mm}

It is proposed in \cite{newhat} that type 0A string in two dimensions 
has a matrix 
model realization 
which is 
a quantum mechanics of a complex matrix. 
Our interest is on type 0A string whose ``time direction'', i.e. the 
direction corresponding 
to the 
free boson, is compactified on $S^1$. 
The compactification is realized simply by letting the time direction of 
the quantum mechanical system 
be periodic. 
As mentioned in \cite{newhat}, the background RR flux can be introduced by 
adding 
a Chern-Simons like term. 
The action of the matrix model is 
\begin{equation}
S = \int_0^{2\pi R}dx\ \mbox{Tr}\left[ |D_xt|^2-\frac1{2\alpha'}|t|^2
+if(A-\tilde{A})\right]. 
\end{equation}
Here $t(x)$ is an $N\times N$ complex matrix field on $S^1$ with radius $R$. 
$t(x)$ couples to the $U(N)$ gauge fields $A,\tilde{A}$ through the covariant derivative 
\begin{equation}
D_xt = \partial_xt-iAt+it\tilde{A}. 
\end{equation}
The constant $f$ is the strength of the RR flux.
If we interpret the matrix model as a quantum mechanics based on D0-branes\cite{Verlinde}, 
the theory 
describes $f$ D0-branes in type 0A theory

Note that there is $i$ in front of the Chern-Simons like term since we have done the Wick 
rotation. 
>From now on, we set $\alpha'=1/2$. 

We fix the gauge by the condition $\partial_xA=\partial_x\tilde{A}=0$. 
Then the remaining degrees of freedom of gauge fields are the zero modes $A_0,\tilde{A}_0$. 
These zero modes can be absorbed by defining a new field $\tilde{t}(x)$ as 
\begin{equation}
\tilde{t}(x) = e^{-iA_0x}t(x)e^{i\tilde{A}_0x}. 
\end{equation}
The new field $\tilde{t}(x)$ obeys a twisted boundary condition 
\begin{equation}
\tilde{t}(x+2\pi R) = U\tilde{t}(x)V^\dag,
\end{equation}
where 
\begin{equation}
U=\exp(-2\pi iRA_0), \hspace{5mm} V=\exp(-2\pi iR\tilde{A}_0). 
\end{equation}

The action for $\tilde{t}$ is 
\begin{equation}
S = \int_0^{2\pi R}dx\ \mbox{Tr}\left[ |\partial_x\tilde{t}|^2-|\tilde{t}|^2 \right]
    -f\mbox{Tr}(\log U-\log V),  
      \label{0Aaction}
\end{equation}
and the partition function is 
\begin{equation}
Z_N = \int {\cal D}U{\cal D}V{\cal D}\tilde{t}(x)\ e^{-S}, 
\end{equation}
where ${\cal D}U, {\cal D}V$ are the Haar measures on $U(N)$. 

One can perturb the action (\ref{0Aaction}) by the Wilson line operators. 
For the gauge choice we chose, the Wilson line operators for the gauge 
fields $A,\tilde{A}$ are 
proportional to Tr$U$, Tr$V$, respectively. 
We will consider an action with a general perturbation  
\begin{equation}
Z_N(\lambda,\tilde{\lambda}) 
= \int {\cal D}U{\cal D}V{\cal D}\tilde{t}(x)\ e^{-S+W(U)+\tilde{W}(V)}, 
\end{equation}
where 
\begin{equation}
W(U) = \sum_{n\ne 0}\lambda_n\mbox{Tr}U^n, \hspace{5mm} 
\tilde{W}(V) = \sum_{n\ne 0}\tilde{\lambda}_n\mbox{Tr}V^n. 
\end{equation}

Note that the above path integral is not well-defined since the potential 
is not bounded from below. 
In the following, we will evaluate 
the path integral by first replacing the unbounded potential with the 
ordinary harmonic potential with frequency $\omega$, and then continuing 
$\omega$ to $i$. 

\vspace{5mm}

The path integral of $\tilde{t}(x)$ can be performed explicitly, since the 
system consists of $N^2$ 
harmonic oscillators with the twisted boundary conditions 
\begin{equation}
\tilde{t}_{ij}(x+2\pi R) = z_i\tilde{z}_j^{-1}\tilde{t}_{ij}(x), 
\end{equation}
where $z_i,\tilde{z}_i$ are eigenvalues of $U,V$, respectively. 
The partition function is, after integration, 
\begin{eqnarray}
Z_N(\lambda,\tilde{\lambda}) 
&=& \frac1{(N!)^2}\oint \prod_i\frac{dz_i}{2\pi iz_i}\frac{d\tilde{z}_i}{2\pi i\tilde{z}_i}
    e^{w(z_i)+\tilde{w}(\tilde{z}_i)} \nonumber \\
& & \times\prod_{i<j}|z_i-z_j|^2|\tilde{z}_i-\tilde{z}_j|^2
    \prod_{i,j}\frac{z_i\tilde{z}_j}{(z_iq^{1/2}-\tilde{z}_jq^{-1/2})
(\tilde{z}_jq^{1/2}-z_iq^{-1/2})},
      \label{Z_N} 
\end{eqnarray}
where $q=\exp(2\pi iR)$ and 
\begin{equation}
w(z) = f\log z+\sum_{n\ne0}\lambda_nz^n, \hspace{5mm} 
\tilde{w}(\tilde{z}) = -f\log z+\sum_{n\ne0}\tilde{\lambda}_n\tilde{z}^n. 
\end{equation}

Comparing with the bosonic case\cite{KKK}, we have two kinds of perturbations 
instead of one. These correspond to the two winding mode perturbations 
in the Liouville theory. In the T-dual 0B theory, these correspond to 
the two kinds of momentum mode perturbations of the Liouville theory
and to the deformations of two Fermi seas in the dual matrix quantum mechanics\cite{TT},
\cite{newhat}.
In later sections, we show some evidences that there is holography between type 0A
matrix model with a suitable deformation without RR flux and ${\cal N}=1$ Witten's 
black hole. We expect that  more general cases with two kinds of winding mode perturbations 
correspond to black hole solutions with two parameters in \cite{Berkovits}. 

Note also that turning on RR flux simply corresponds to turning on $\log z$ 
perturbation. It might be interesting to find integrable structures associated with 
the backgrounds with RR flux. In the appendix \ref{CFT} we show that the partition function 
admits a simple free field realization which generalize the result at \cite{Yin} where 
the free field realization for the case without RR flux was discussed.
There we see that turning on RR flux corresponds to nonzero relative momentum 
of the two free bosons appearing in the free field realization. 
In the appendix \ref{CFT} we derive the analogue of the Hirota bilinear differential 
equation for the model. However the resulting equation is more complicated 
and we do not reduce the equation into a known form of $\tau$ functions.
In the next section we show that the equation could be explicitly evaluated 
to give the product of the $\tau$ function when two winding modes are the same,
which dictates that there should be no RR flux.
Unlike the discussion in \cite{Yin}, this is the exact expression including 
nonperturbative corrections. Thus it might be interesting to work out 
nonperturbative terms and to compare the expression with the corresponding computations 
in the Liouville theory as is done in\cite{Alexandrov}.

\vspace{1cm}

\section{Determinant representation and factorization}

\vspace{5mm}

The expression (\ref{Z_N}) for the partition function can also be rewritten as 
\begin{eqnarray}
Z_N(\lambda,\tilde{\lambda}) 
&=& \frac1{(N!)^2}\oint\prod_{i=1}^N\frac{dz_i}{2\pi i}\frac{d\tilde{z}_i}{2\pi i}
    e^{w(z_i)+\tilde{w}(\tilde{z}_i)}
    \mbox{det}_{ij}\left( \frac1{z_iq^{1/2}-\tilde{z}_jq^{-1/2}} \right)
    \mbox{det}_{ij}\left( \frac1{\tilde{z}_jq^{1/2}-z_iq^{-1/2}} \right) \nonumber \\
&=& \frac1{(N!)^2}\oint\prod_{i=1}^N\frac{dz_i}{2\pi i}\frac{d\tilde{z}_i}{2\pi i}
    \mbox{det}_{ij}\left( \frac{e^{(w(z_i)+\tilde{w}(\tilde{z}_j))/2}}{z_iq^{1/2}
-\tilde{z}_jq^{-1/2}} 
    \right)
    \mbox{det}_{ij}\left( \frac{e^{(\tilde{w}(\tilde{z}_j)+w(z_i))/2}}{\tilde{z}_jq^{1/2}
-z_iq^{-1/2}} 
    \right), 
       \label{detdet}
\end{eqnarray}
by using the identity 
\begin{equation}
\frac{\Delta(a)\Delta(b)}{\prod_{ij}(a_i-b_j)} = 
\mbox{det}_{ij}\left( \frac1{a_i-b_j} \right), 
\end{equation}
where $\Delta(a)$ is the Vandermonde determinant. 

As in the bosonic case, it is more convenient to consider the grand canonical 
partition function 
$Z(\lambda,\tilde{\lambda};\mu)$ defined as follows, 
\begin{equation}
Z(\lambda,\tilde{\lambda};\mu) = \sum_{N=0}^\infty 
e^{2\pi R\mu N}Z_N(\lambda,\tilde{\lambda}). 
   \label{GCPF}
\end{equation}
This partition function has the following determinant representation 
\begin{equation}
Z(\lambda,\tilde{\lambda};\mu) = \mbox{Det}(1+e^{2\pi R\mu}{\cal K}), 
\end{equation}
where 
\begin{eqnarray}
{\cal K}(z,z') &=& \oint\frac{dz''}{2\pi i}K(z,z'')\tilde{K}(z'',z'), \\
K(z,z') &=& \frac{e^{(w(z)+\tilde{w}(z'))/2}}{zq^{1/2}-z'q^{-1/2}}, \\
\tilde{K}(z,z') &=& \frac{e^{(\tilde{w}(z)+w(z'))/2}}{zq^{1/2}-z'q^{-1/2}}. 
\end{eqnarray}
The calculation is summarized in appendix \ref{determinant}. 

When $K=\tilde{K}$, the determinant is factorized into two factors, 
\begin{equation}
Z(\lambda,\tilde{\lambda};\mu) 
= \mbox{Det}(1+ie^{\pi R\mu}K)\cdot\mbox{Det}(1-ie^{\pi R\mu}\tilde{K}). 
   \label{factor}
\end{equation}
Note that in this case the RR flux $f$ must vanish and 
$\lambda_n=\tilde{\lambda}_n$. 
Remarkably, each determinant factor is the 
partition function investigated in \cite{KKK} with the chemical potential replaced with 
$(\mu\pm i/2R)/2$. 
Therefore, $Z(\lambda,\tilde{\lambda};\mu)$ can be obtained by applying the same technique 
used in 
\cite{KKK}. 

When all $\lambda,\tilde{\lambda}$'s vanish, the determinants can be calculated explicitly. 
One can show that
\footnote{These calculations are done with $q=\exp(2\pi\omega R)>1$ and $\omega$ is analytically 
continued after the contour integrations.} 
\begin{eqnarray}
K\cdot z^n 
&=& \oint\frac{dz'}{2\pi i}K(z,z')(z')^n \nonumber \\
&=& \left\{
\begin{array}{lc}
0, & (n\ge f/2), \\ q^{n+1/2-f/2}z^n, & (n<f/2), 
\end{array}
\right. \\
\tilde{K}\cdot z^n 
&=& \oint\frac{dz'}{2\pi i}\tilde{K}(z,z')(z')^n \nonumber \\
&=& \left\{
\begin{array}{lc}
0, & (n\ge -f/2), \\ q^{n+1/2+f/2}z^n, & (n<-f/2), 
\end{array}
\right. \\
{\cal K}\cdot z^n 
&=& \oint\frac{dz'}{2\pi i}{\cal K}(z,z')(z')^n \nonumber \\
&=& \left\{
\begin{array}{lc}
0, & (n\ge -|f|/2), \\ q^{2n+1}z^n, & (n<-|f|/2). 
\end{array}
\right. 
\end{eqnarray}
We have assumed that $f/2$ is an integer. 
Therefore one obtains 
\begin{eqnarray}
\mbox{Det}(1+e^{2\pi R\mu}{\cal K}) 
&=& \prod_{n<-|f|/2}(1+ie^{\pi R\mu}q^{n+1/2})(1-ie^{\pi R\mu}q^{n+1/2}) \nonumber \\
&\ne& \mbox{Det}(1+ie^{\pi R\mu}K)\cdot\mbox{Det}(1-ie^{\pi R\mu}\tilde{K}). 
\end{eqnarray}
This implies that the grand canonical 
partition function is not factorized into two determinants in a similar way as in 
(\ref{factor}) 
when the 
RR flux is turned on.

\vspace{1cm}

\section{Free energy of 0A matrix model with perturbation}

\vspace{5mm}

\subsection{Review of bosonic case}

\vspace{5mm}

In \cite{KKK}, the c=1 matrix model with a perturbation is investigated and its free energy 
is calculated 
perturbatively. 
The grand canonical partition function of this system is defined similarly to (\ref{GCPF}), and 
this partition function also has the following determinant representation, 
\begin{equation}
Z(\lambda,\mu) = \mbox{Det}(1+e^{2\pi R\mu}K), 
   \label{bosonicdet}
\end{equation}
where 
\begin{equation}
K(z,z') = \frac{e^{u(z)+u(z')}}{zq^{1/2}-z'q^{-1/2}}, \hspace{5mm} 
u(z) = \frac12\sum_{n\ne0}\lambda_nz^n. 
\end{equation}
It is shown in \cite{KKK} 
that when all $\lambda$'s except for $\lambda_{\pm1}$ vanish corresponding to the 
Sine-Liouville theory, the partition function 
$Z(\lambda,\mu)$ is obtained by solving a non-linear differential equation. 
For $\chi(\lambda,\mu)=\partial_\mu^2F(\lambda,\mu)$, where $F(\lambda,\mu)=\log Z(\lambda,\mu)$, 
the differential equation is 
\begin{equation}
\frac1{4\lambda}\partial_\lambda(\lambda\partial_\lambda\chi(\lambda,\mu)) 
+\partial_\mu^2\exp\left[ -\left( \frac{\sin(\partial_\mu/2)}{\partial_\mu/2} \right)^2
\chi(\lambda,\mu) \right] = 0, 
   \label{diff}
\end{equation}
where $\lambda\propto\sqrt{\lambda_{+1}\lambda_{-1}}$ and 
\begin{equation}
\frac{\sin(\partial_\mu/2)}{\partial_\mu/2} 
= \sum_{n=0}^\infty (-1)^n\frac{2^{-2n}}{(2n+1)!}\partial_\mu^{2n}. 
\end{equation}
$F(\lambda=0,\mu)$ can be calculated explicitly, 
\begin{eqnarray}
\partial_\mu^3F(\lambda=0,\mu) 
&=& \partial_\mu^3\prod_{n=0}^\infty\log(1+e^{2\pi R\mu}q^{-n-1/2}) \nonumber \\
&=& R\ \mbox{Im}\int_0^\infty dt\ e^{-it(\mu-i\epsilon)}
    \frac{t/2}{\sinh(t/2)}\frac{t/2R}{\sinh(t/2R)}, 
\end{eqnarray}
where $\epsilon=+0$, 
and this quantity can be used as a boundary condition 
to the solution of the differential equation (\ref{diff}). 

To solve the equation (\ref{diff}), $\chi(\lambda,\mu)$ is assumed to have 
a perturbative expansion 
whose expansion parameter is a power of $\lambda$ and the coefficient 
at each order is a function 
of $y\propto\frac{\mu}{\lambda^{2/(2-R)}}$. 
Substituting this perturbative series into (\ref{diff}), the differential equation is reduced to 
an infinite number of second order differential equations. 
In \cite{KKK}, they are solved explicitly up to genus one. 

\vspace{5mm}

\subsection{Type 0A case}

\vspace{5mm}

We would like to evaluate the determinant 
\begin{equation}
Z_\pm(\lambda,\mu) = \mbox{Det}(1\pm ie^{\pi R\mu}K)|_{\lambda=\tilde{\lambda}}, 
\end{equation}
from which the grand canonical partition function of type 0A matrix model is obtained as 
\begin{equation}
Z(\lambda,\tilde{\lambda}=\lambda;\mu) = Z_+(\lambda,\mu)Z_-(\lambda,\mu). 
\end{equation}
One can see that $Z_\pm(\lambda,\mu)$ is the same function as the bosonic partition function 
(\ref{bosonicdet}) with $\mu$ replaced with $(\mu\pm i/2R)/2$. 
Therefore, when all $\lambda$'s except for $\lambda_{\pm1}$ are turned off, 
$Z_\pm(\lambda,\mu)$ can also be determined by solving the differential equation 
\begin{equation}
\frac1{4\lambda}\partial_\lambda(\lambda\partial_\lambda\chi_\pm(\lambda,\mu')) 
+\partial_{\mu'}^2\exp\left[ -\left( \frac{\sin(\partial_{\mu'}/2)}{\partial_{\mu'}/2} \right)^2
\chi_\pm(\lambda,\mu') \right] = 0, 
   \label{diff'}
\end{equation}
where $\mu'=\mu/2$ and $\chi_\pm(\lambda,\mu)=\partial_{\mu'}^2\log Z_\pm(\lambda,\mu)$. 

$F_\pm(\lambda=0,\mu)$ can be calculated explicitly as follows, 
\begin{eqnarray}
\partial_{\mu'}^3F_\pm(\lambda=0,\mu) 
&=& \frac R{i}\int_0^\infty dt\ e^{-it(\mu'\pm i/4R-i\epsilon)}
    \frac{t/2}{\sinh(t/2)}\frac{t/2R}{\sinh(t/2R)} \nonumber \\
& &-\frac R{i}\int_0^\infty dt\ e^{+it(\mu'\pm i/4R+i\epsilon)}
    \frac{t/2}{\sinh(t/2)}\frac{t/2R}{\sinh(t/2R)} \nonumber \\
&=&-2R\int_0^\infty dt\ \sin((\mu'\pm i/4R)t)
    \frac{t/2}{\sinh(t/2)}\frac{t/2R}{\sinh(t/2R)}  \nonumber \\
&=& 2R\ \mbox{Im}\int_0^\infty dt\ e^{-it(\mu'-i\epsilon)}
    \frac{t/2}{\sinh(t/2)}\frac{t/4R}{\sinh(t/4R)} \nonumber \\
& &\mp2iR\ \mbox{Re}\int_0^\infty dt\ e^{-it(\mu'-i\epsilon)}
    \frac{t/2}{\sinh(t/2)}\frac{t/4R}{\cosh(t/4R)}. 
\end{eqnarray}
>From this expression, one can obtain the asymptotic expansion of $\chi_\pm(\lambda=0,\mu)$, 
\begin{eqnarray}
\chi_\pm(\lambda=0,\mu) 
&\sim& -2R\log\mu'+2R\sum_{N=1}^\infty f_N(2R)\mu'^{-2N} \nonumber \\
& & \mp2iR\sum_{N=0}^\infty g_N(2R)\mu'^{-2N-1}, 
   \label{asymptotic}
\end{eqnarray}
where 
\begin{eqnarray}
f_N(2R) &=& (2N-1)!\sum_{k=0}^N\frac{|2^{2N-2k}-2||2^{2k}-2|}{(2N-2k)!(2k)!}|B_{2N-2k}B_{2k}|
            (2R)^{-2k}2^{-2N}, \nonumber \\
g_N(2R) &=& (2N)!\sum_{k=0}^N\frac{|2^{2N-2k}-2|}{(2N-2k)!(2k)!}|B_{2N-2k}E_{2k}|
(2R)^{-2k-1}2^{-2N-1}. 
\end{eqnarray}
Here $B_{k},E_{k}$ are the Bernoulli number and the Euler number, respectively, 
\begin{eqnarray}
\frac x{e^x-1} &=& \sum_{k=0}^\infty \frac{B_k}{k!}x^k, \\
\mbox{sech}\ x &=& \sum_{k=0}\frac{E_{k}}{k!}x^{k}. 
\end{eqnarray}
The asymptotic expansion (\ref{asymptotic}) will be used as a boundary condition 
in solving the 
differential equation. 

We make an ansatz for $\chi_\pm(\lambda,\mu)$ as follows, 
\begin{eqnarray}
\chi_\pm(\lambda,\mu) &=& -2R\log\mu'+\sum_{n=0}^\infty \mu'^{-n}\chi_{\pm,n}(z), 
\end{eqnarray}
where $\chi_{\pm,n}(z)$ is a power series in $z=\lambda^2/\mu'^{2-2R}$, following \cite{KKK}, 
and obeys the 
boundary condition 
\begin{eqnarray}
\chi_{\pm,0}(0) &=& 0, \nonumber \\
\chi_{\pm,2n}(0) &=& 2Rf_n(2R), \nonumber \\
\chi_{\pm,2n+1}(0) &=& \mp2iRg_n(2R). 
\end{eqnarray}
According to \cite{KKK}, we make a change of variables from $(\mu',z)$ to $(\lambda,y)$ where 
\begin{equation}
y = z^{-1/(2-2R)} = \frac{\mu'}{\lambda^{1/(1-R)}}. 
\end{equation}
Then the result is 
\begin{eqnarray}
\chi_\pm(\lambda,\mu) &=& -\frac{2R}{1-R}\log\lambda
                          +\sum_{n=0}^\infty\lambda^{-n/(1-R)}X_{\pm,n}(y), \nonumber \\
X_{\pm,0}(y) &=& -2R\log y+\chi_{\pm,0}(y^{-1/(2-2R)}), \label{genus0bc} \nonumber \\
X_{\pm,n}(y) &=& y^{-n}\chi_{\pm,n}(y^{-1/(2-2R)}). \hspace{5mm} (n\ge1)  \label{generalbc}
\end{eqnarray}
Substituting this ansatz into (\ref{diff'}) and setting to zero 
the coefficient of each power of $\lambda$, 
we obtain the following set of equations, 
\begin{eqnarray}
(y\partial_y+n)^2X_{\pm,n}(y) 
+ 4(1-R)^2\partial_y^2\left( e^{-X_{\pm,0}(y)}p_n[X_\pm] \right) = 0, 
   \label{perturbativeEq}
\end{eqnarray}
where $p_n[X_\pm]$ is defined as follows, 
\begin{equation}
\exp\left[ -\sum_{n=0}^\infty x^n\left( \frac{\sin(x\partial_y/2)}{x\partial_y/2} \right)^2
X_{\pm,n}(y) \right] 
= \sum_{n=0}^\infty x^ne^{-X_{\pm,0}(y)}p_n[X_\pm]. 
\end{equation}
Explicit forms of $p_n[X_\pm]$ are 
\begin{eqnarray}
p_0[X_\pm] &=& 1, \nonumber\\
p_1[X_\pm] &=& -X_{\pm,1}, \nonumber \\
p_2[X_\pm] &=& -X_{\pm,2}+\frac1{12}\partial_y^2X_{\pm,0}+\frac12(X_{\pm,1})^2, \nonumber\\
p_3[X_\pm] &=& -X_{\pm,3}+\frac1{12}\partial_y^2X_{\pm,1}
-\frac1{12}X_{\pm,1}\partial_y^2X_{\pm,0}
               +X_{\pm,1}X_{\pm,2}-\frac16(X_{\pm,1})^3, \nonumber\\
p_4[X_\pm] &=& -X_{\pm,4}+\frac1{12}\partial_y^2X_{\pm,2}-\frac1{360}\partial_y^4X_{\pm,0}
               -\frac1{12}X_{\pm,1}\partial_y^2X_{\pm,1}+X_{\pm,1}X_{\pm,3} \nonumber \\
           & & +\frac1{288}(\partial_y^2X_{\pm,0})^2-\frac1{12}\partial_y^2X_{\pm,0}X_{\pm,2}
               +\frac12(X_{\pm,2})^2 \nonumber \\
           & & +\frac1{24}(X_{\pm,1})^2\partial_y^2X_{\pm,0}-\frac12(X_{\pm,1})^2X_{\pm,2}
               +\frac1{24}(X_{\pm,1})^4, \\               
           & & \mbox{etc.} \nonumber 
\end{eqnarray}

\vspace{3mm}

{\it (i) $n=0$}

\vspace{3mm}

The differential equation which we have to solve is 
\begin{equation}
(y\partial_y)^2X_{\pm,0}(y)+4(1-R)^2\partial_y^2e^{-X_{\pm,0}(y)} = 0. 
\end{equation}
In \cite{KKK}, this equation is solved and the solution is 
\begin{equation}
y=e^{-X_{\pm,0}/2R}-(2R-1)e^{(1-2R)X_{\pm,0}/2R}, 
\end{equation}
which is valid for $0<R<1$. 
For $R>1$, there is no solution satisfying the boundary condition (\ref{genus0bc}). 
When $R=1/2$, which would be relevant to the fermionic cigar geometry, the solution has 
the simple 
form 
\begin{equation}
X_{\pm,0} = -\log y. 
\end{equation}
Then the tree level contribution to $F_\pm(\lambda,\mu)$ is 
\begin{eqnarray}
F_\pm(\lambda,\mu)|_{tree} 
&=& -\frac12(\mu')^2\log\mu'+A_1\mu'\lambda^2+A_2\lambda^4 \nonumber \\
&=& -\frac12(\mu')^2\log\mu'+\mu'^2(A_1z+A_2z^2), 
\end{eqnarray}
where $z=\lambda^2/\mu'$. 
We have omitted terms which depend only on $\mu'$. 
The constants $A_1, A_2$ can be fixed by matching this expression with the (3.7) of 
\cite{KKK} with $R=1$, 
and the result is $A_1=-1, A_2=0$.

\vspace{3mm}

{\it (ii) $n\ge1$}

\vspace{3mm}

From now on, we focus on the case $R=1/2$. 

One can see that $p_n[X_\pm]$ has the structure 
\begin{equation}
p_n[X_\pm] = -X_{\pm,n}+\tilde{p}_n[X_\pm], 
\end{equation}
where $\tilde{p}_n[X_\pm]$ only contains $X_{\pm,m}$ with $m<n$. 
Then the equation (\ref{perturbativeEq}) is 
\begin{equation}
y(1-y)\partial_y^2X_{\pm,n}+(2-(2n+1)y)\partial_yX_{\pm,n}-n^2X_{\pm,n} 
= \partial_y^2\left( y\tilde{p}_n[X_\pm] \right). 
\end{equation}
This equation can be solved order by order in $n$. 
It is remarkable that the equation LHS$=0$ is the hypergeometric equation, and therefore the 
general solution would have singularities at $y=0,1$. 

The solutions which satisfy the boundary conditions (\ref{generalbc}) are as follows, 
\begin{eqnarray}
X_{\pm,1} &=& \mp\frac i{2y}, \nonumber \\
X_{\pm,2} &=& -\frac1{24y^2}+\frac1{8(y-1)^2}, \nonumber \\
X_{\pm,3} &=& \mp i\left( \frac1{24y^3}+\frac1{6(y-1)^3}+\frac1{8(y-1)^4} \right), \nonumber\\
X_{\pm,4} &=& -\frac7{320y^4}+\frac3{64(y-1)^4}+\frac{17}{32(y-1)^5}+\frac5{8(y-1)^6}, \\
          & & \mbox{etc.} \nonumber
\end{eqnarray}
These results suggest that for general $n$ the solution $X_{\pm,n}$ would have the following 
simple 
form 
\begin{equation}
X_{\pm,n} = \frac{c}{y^n}+\sum_{k=n}^{2n-2}\frac{c_k}{(y-1)^k}. 
   \label{conjecture}
\end{equation}
where $c,c_k$ are rational numbers. 

Now $F_\pm(\lambda,\mu)$ can be obtained by integrating twice, and we obtain 
with $z=\frac{1}{y}$ when $R=\frac{1}{2}$ and $y=\frac{\mu '}{\lambda^2}$
\begin{eqnarray}
F_{\pm,1}(\lambda,\mu) &=& \mp\frac i2\mu'\log\mu'+a_1^\pm\mu'z, \\
F_{\pm,2}(\lambda,\mu) &=& -\frac1{12}\log\mu'-\frac18\log(1-z), 
   \label{oneloop} \\
F_{\pm,3}(\lambda,\mu) &=& \mp\frac i{\mu'}
                           \left( \frac1{48}+\frac1{12(1-z)}+\frac z{48(1-z)^2} \right), \\
F_{\pm,4}(\lambda,\mu) &=& \frac1{\mu'^2}\left( -\frac7{1920}+\frac1{128(1-z)^2}
                          +\frac{17z}{384(1-z)^3}+\frac{z^2}{32(1-z)^4} \right), \\
                       & & \mbox{etc.} \nonumber 
\end{eqnarray}

Note that the terms depending only on $\mu '$ in $F_{\pm, 0}, F_{\pm, 2}$
are a half of the corresponding part of the usual genus expansion of the type 0A matrix model.
Also all of $F_{+, 2n+1}$ terms are cancelled against $ F_{-, 2n+1}$ in the final 
expression of the free energy. 

Note that we have two singularities in the expansion at $y=0$ and $y=1$.
The singularities at $y=0$ corresponds to the limit 
$\mu' \rightarrow 0$. This singularity appears even when we turn off the 
deformations of the type 0A matrix model. For the bosonic case, this is interpreted 
as the vanishing of the mass gap of the corresponding conformal field theory 
on the Witten's black hole background and the same interpretation could be 
applied to the type 0A case. If we use the relation between the c=1 matrix 
model at the self-dual radius and the topological string on the conifold, 
this singularity at $y=0$ corresponds to the conifold singularity of the topological
string. The vanishing of the mass gap is translated into the appearance of the 
massless black hole at the conifold point. Thus new states are added to the string 
perturbation theory near the conifold. The singularities at $y=1$ is hitherto unknown.
>From the Wilson line computation in the next section, we propose that this 
corresponds to the point of the enhancement of the worldsheet supersymmetry 
from ${\cal N}=1$ to ${\cal N}=2$. It is desirable to collect further evidences for this 
proposal.

\vspace{5mm}

Similar calculations can be done for the bosonic case with $R=1$ and 
type 0B case with $R=1/2$. Type 0B case has the same differential equation 
as the bosonic case but 
with the different boundary conditions, that is, the free energy of type 0B without perturbation is 
twice of that of bosonic string. 

Firstly, consider the bosonic case. 
The boundary condition at $\lambda=0$ is 
\begin{equation}
\chi_b(\lambda=0,\mu) 
\sim -\log\mu+\sum_{N=1}^\infty f_N(1)\mu^{-2N}. 
\end{equation}
We make an ansatz 
\begin{eqnarray}
\chi_b(\lambda,\mu) &=& -\log\mu+\sum_{n=0}^\infty \mu^{-2n}\chi_{b,n}(z), 
\end{eqnarray}
where $\chi_{b,n}(z)$ is a power series in $z=\lambda^2/\mu$, following \cite{KKK}, 
and obeys the 
boundary condition 
\begin{eqnarray}
\chi_{b,0}(0) &=& 0, \nonumber \\
\chi_{b,n}(0) &=& f_n(1).
\end{eqnarray}
Through almost the same procedure, we obtain for the bosonic case, 
\begin{eqnarray}
X_{b,0} &=& -\log y, \nonumber \\
X_{b,1} &=& \frac1{12y^2}, \nonumber \\
X_{b,2} &=& \frac1{40y^4}, \nonumber \\
X_{b,3} &=& \frac5{252y^6}, \\
    & & \mbox{etc.} \nonumber 
\end{eqnarray}
Here $X_{b,n}(y)$ is defined as follows,
\begin{eqnarray}
X_{b,0}(y) &=& -\log y+\chi_{b,0}(y^{-1/(2-R)}), \nonumber \\
X_{b,n}(y) &=& y^{-2n}\chi_{b,n}(y^{-1/(2-2R)}). \hspace{5mm} (n\ge1) 
\end{eqnarray}
It is tempting to conjecture that 
\begin{equation}
X_{b,n} = \frac{f_n(1)}{y^{2n}} \hspace{5mm} (n\ge1), 
\end{equation}
which implies that $\chi_b(\lambda,\mu)$ is {\it independent} of $\lambda$. 
In fact, it can be proved that this is the case, see appendix \ref{proof}. 

For type 0B case, the boundary condition is 
\begin{equation}
\chi_B(\lambda=0,\mu) = 2\chi_b(\lambda=0,\mu), 
\end{equation}
but now $R=1/2$. 
That is, 
\begin{equation}
\chi_{B,0}(0) = 0, \hspace{5mm} \chi_{B,n\ge1}(0) = f_n(1/2). 
\end{equation}
The form of $X_{B,n}$ is not $2X_{b,n}$ since the equation is non-linear, and we obtain 
\begin{eqnarray}
X_{B,0} &=& -\log y, \nonumber \\
X_{B,1} &=& \frac1{12y^2}+\frac1{8(y-1)^2}, \nonumber \\
X_{B,2} &=& \frac1{40y^4}+\frac9{64(y-1)^4}-\frac{9}{32(y-1)^5}, \nonumber \\
X_{B,3} &=& \frac5{252y^6}+\frac{45}{128(y-1)^6}+\frac{1215}{512(y-1)^7}
+\frac{1323}{512(y-1)^8}, \\
        & & \mbox{etc.} \nonumber 
\end{eqnarray}
It looks that the type 0B result is simpler than the type 0A result. 
Note that $X_{B,n}$ corresponds to $X_{\pm,2n}$ in the type 0A case.

\vspace{1cm}

\section{Wilson line}

\vspace{5mm}

In \cite{SY}, it is shown  that the vev of Wilson line operator 
along the (Euclideanized) 
time direction provides the correct information on the target space geometry, 
assuming a holographic 
relation. It is interesting to check whether a similar calculation can be done 
in the cigar geometry 
in type 0A string. 

It is expected that the vev of the Wilson line is related to a holographically dual string theory by 
the following relation, 
\begin{equation}
\langle W \rangle = \langle e^{i\oint A} \rangle = \int DX\ e^{-S_p(X)}, 
\end{equation}
where $S_p(X)$ is the Polyakov action. 
There is another relation which uses the Nambu-Goto action, 
\begin{equation}
\left\langle\frac1N e^{i\oint A} \right\rangle = \int DX\ e^{-S_{NG}}
\end{equation}
as is used in \cite{SY}. 
These two relations would be different if the two string theories are different at the quantum level. 
At least we can say that the former has an advantage since the string coupling dependence is manifest. 
In the following, we will use the former. 

The leading order behavior comes from the classical value of the action. 
Since the worldsheet of the string wraps all the spacetime, the worldsheet coordinates are identified 
with the spacetime ones, and the worldsheet metric is also the same as the spacetime one. 
Then, 
\begin{eqnarray}
S_p 
&=& \frac1{2\pi \alpha'}\int d^2x\sqrt{g}+\Phi(0)-\frac1{4\pi}\int d^2x\ R^{(2)}\log(\cosh\ r) 
    \nonumber \\
&=& k(\Phi(0)-\Phi_b)+\Phi(0)+O(1), 
\end{eqnarray}
where $\Phi_b$ is the value of the dilaton at the point where the boundary theory lives. 
That is, the Wilson line behaves as 
\begin{equation}
\langle W \rangle \sim e^{-3\Phi(0)/2}. 
\end{equation}

Let us compare this result with a quantity derived from the matrix model. 
The vev of the Wilson line can be calculated as follows, 
\begin{equation}
\langle W \rangle_{mm} = \langle \mbox{Tr}U \rangle \sim \partial_\lambda F(\lambda,\mu'). 
\end{equation}
Note that the vev of Tr$V$ is the same since the perturbation is symmetric. 

If we consider a limit $\lambda\to\infty$ while $\mu'$ is fixed, in analogy with 
the bosonic case\cite{SY}, 
we obtain 
\begin{equation}
\langle W \rangle_{mm} \sim \lambda \sim g_s^{-1}, 
\end{equation}
so this limit is not appropriate for the fermionic case. 
The correct behavior $\langle W \rangle_{mm}\sim g_s^{-3/2}$ of the Wilson line is obtained by taking 
another limit $\lambda,\mu'\to\infty$ while $y$ is fixed. 
Therefore, the type 0A matrix model with the symmetric perturbation 
can be holographically dual to the type 0A string on the Witten's 
black hole by taking the latter limit. 

One interesting fact is that ${\cal N}=1 SL(2, R)/U(1)$ supercoset theory 
has an accidental ${\cal N}=2$ worldsheet supersymmetry \cite{KKK}.
Furthermore in \cite{HK}, it is shown that the ${\cal N}=1$ supercoset theory 
is mirror to ${\cal N}=2$ Liouville theory. Thus from the holographic point of view 
it is natural  that the  limit of the deformed matrix model
$\lambda, \mu' \rightarrow \infty$ with 
$y$ fixed  corresponds to the point of enhanced 
${\cal N}=2$ worldsheet supersymmetry. This is an interesting matrix model 
realization of ${\cal N}=2$ Liouville theory and we should look for further 
evidences for this holographic relation. Note also that in terms of ${\cal N}=1$
superfield language, we are taking the limit where the coefficient of 
${\cal N}=1$ Liouville potential terms are vanishing while keeping the 
coefficient of ${\cal N}=1$ sine-Liouville terms finite. 

Note that we have considered the situation with $y=O(1)$. 
This seems to be justified by the singular behavior of the free energy at $y=1$. 
The $\log(1-z)$ term in the genus one free energy (\ref{oneloop}) 
would imply that a massless state appears when 
we take $y\to1$ limit. 
This seems to be consistent with the fact that, in the worldsheet theory, there is an 
IR fixed point 
at which the supersymmetry is enhanced to ${\cal N}=2$. 
Note that at some general $R$, the structure of the tree level free energy is almost 
the same as 
that of the bosonic one except 
for a scaling of $R$, which would suggest that the type 0A 
matrix model with generic $R$ describes ${\cal N}=1$ 
Liouville theory with a perturbation as also claimed in \cite{KKK}. 
According to this interpretation of the singularity at $y=1$, since this 
is an IR divergence, the 
singularity does not mean any inconsistency of the model, and therefore the above 
calculation of 
Wilson line around $y=1$ could be justified. 
Note that the higher order contribution to the Wilson line can be finite and non-zero 
by fine-tuning 
how fast $y$ goes to 1, if the conjectured form (\ref{conjecture}) is correct. 
The above calculation suggests that in order to use the deformed matrix model 
to describe ${\cal N}=2$ Liouville theory properly, we should take account of additional 
degrees of freedom appearing in the limit of $y\rightarrow 1$.
If we use the relation of the deformed matrix model to the topological string 
as explained in the next section, we can gain some insights for the additional 
degrees of freedom to be included.

\vspace{1cm}

\section{Relation to topological string}

\vspace{5mm}


The  free energy of type 0A matrix model is 
\begin{equation}
\partial_\mu^3 F_{0A} = R\ \mbox{Im}\int_0^\infty dt\ 
e^{-it(\mu-if-i\epsilon)}\frac{t}{\sinh(t)}
                      \frac{t/2R}{\sinh(t/2R)}, 
\end{equation}
while that of the bosonic string is 
\begin{equation}
\partial_\mu^3 F_{b} = R\ \mbox{Im}\int_0^\infty dt\ 
e^{-it(\mu-i\epsilon)}\frac{t/2}{\sinh(t/2)}
                      \frac{t/2R}{\sinh(t/2R)}. 
\end{equation}

The bosonic case with $R=1$ produces the partition function of type IIB 
topological string on the conifold. The type 0A case with  $R=\frac{1}{2}$ and $f=0$ 
gives the twice of the  partition function of the topological string 
on the conifold. Thus one might wonder if the theory is again related to
a topological string. Indeed it is argued in \cite{Oz} that corresponding 
topological string is defined on a $Z_2$ orbifold of the conifold.
The $Z_2$ orbifold action on the conifold $x'y'=uv$ is given by
\begin{eqnarray}
(x',y') &\rightarrow& (-x',-y')  \nonumber \\
(u,v) & \rightarrow & (u,v).
\end{eqnarray}
If we choose invariant variables 
$x\equiv x'^2, \,\,  y\equiv y'^2$, we have 
\begin{equation}
xy=(uv)^2
\end{equation}
The deformation is given by 
\begin{equation}
xy=(uv+\mu)^2-\frac{f^2}{2}  \label{dc}
\end{equation}
where $f$ is related to the RR flux of type 0A model.
It is explained in \cite{Oz} that the geometry given by eq. (\ref{dc})
has two $S^3$ cycles, which are of size $\mu$ if $f=0$. Thus in the 
$\mu\rightarrow 0$ limit, two $S^3$s are shrinking simultaneously. 
For $y=1$ case of the previous section, this is related to deforming 
the complex structures at infinities of the non-compact Calabi-Yau manifold 
with conifold-like singularities\cite{hierarchy}. In our case, we have 
the same deformations
for both winding modes and this should be related to the deformation 
of the curve $uv+\mu=0$. This would be quite close to the bosonic case
and equivalent picture is also developed in \cite{Timedependent}.
There the complex structure deformation is mapped to the deformation 
of the Fermi sea profile.
It would be interesting to understand the integrable structures of the 
topological string on such $Z_2$ orbifold of the conifold related to the 
complex structure deformations at the asymptotic regions. 

\newpage

\appendix

\vspace{1cm}

\section{Determinant representation} \label{determinant}

\vspace{5mm}

The expression (\ref{detdet}) can be written as follows, 
\begin{eqnarray}
Z_N(\lambda,\tilde{\lambda}) 
&=& \frac1{(N!)^2}\oint\prod_{i=1}^N\frac{dz_i}{2\pi i}\frac{d\tilde{z}_i}{2\pi i}\ 
    \mbox{det}_{ij}K(z_i,\tilde{z}_j)\ \mbox{det}_{ij}\tilde{K}(\tilde{z}_i,z_j) \nonumber \\
&=& \frac1{(N!)^2}\oint\prod_{i=1}^N\frac{dz_i}{2\pi i}\frac{d\tilde{z}_i}{2\pi i}\ 
    \sum_{\sigma\in S_N}(-1)^\sigma \prod_{k=1}^N K(z_k,\tilde{z}_{\sigma(k)})
    \sum_{\sigma'\in S_N}(-1)^{\sigma'}\prod_{l=1}^N \tilde{K}(\tilde{z}_l,z_{\sigma'(l)}) \nonumber \\
&=& \frac1{(N!)^2}\oint\prod_{i=1}^N\frac{dz_i}{2\pi i}\frac{d\tilde{z}_i}{2\pi i}\ 
    \sum_{\sigma\in S_N}(-1)^\sigma \prod_{k=1}^N K(z_k,\tilde{z}_{\sigma(k)})
    \sum_{\sigma'\in S_N}(-1)^{\sigma'}\prod_{l=1}^N 
    \tilde{K}(\tilde{z}_{\sigma(l)},z_{\sigma(\sigma'(l))}) \nonumber \\
&=& \frac1{(N!)^2}\oint\prod_{i=1}^N\frac{dz_i}{2\pi i}\frac{d\tilde{z}_i}{2\pi i}\ 
    \sum_{\sigma\in S_N}(-1)^\sigma \prod_{k=1}^N K(z_k,\tilde{z}_{\sigma(k)})
    \sum_{\tilde{\sigma}\in S_N}(-1)^{\tilde{\sigma}}(-1)^{\sigma}\prod_{l=1}^N 
    \tilde{K}(\tilde{z}_{\sigma(l)},z_{\tilde{\sigma}(l)}) \nonumber \\
&=& \frac1{(N!)^2}\oint\prod_{i=1}^N\frac{dz_i}{2\pi i}\frac{d\tilde{z}_i}{2\pi i}\ 
    \sum_{\sigma\in S_N}\sum_{\tilde{\sigma}\in S_N}(-1)^{\tilde{\sigma}}
    \prod_{k=1}^N K(z_k,\tilde{z}_{\sigma(k)})\tilde{K}(\tilde{z}_{\sigma(k)},z_{\tilde{\sigma}(k)}) 
    \nonumber \\
&=& \frac1{N!}\oint\prod_{i=1}^N\frac{dz_i}{2\pi i}\ 
    \sum_{\tilde{\sigma}\in S_N}(-1)^{\tilde{\sigma}}
    \prod_{k=1}^N \oint\frac{d\tilde{z}}{2\pi i}
    K(z_k,\tilde{z})\tilde{K}(\tilde{z},z_{\tilde{\sigma}(k)}) \nonumber \\
&=& \frac1{N!}\oint\prod_{i=1}^N\frac{dz_i}{2\pi i}\ 
    \sum_{\tilde{\sigma}\in S_N}(-1)^{\tilde{\sigma}}
    \prod_{k=1}^N {\cal K}(z_k,z_{\tilde{\sigma}(k)}). 
\end{eqnarray}
The summation over permutations can be rewritten as follows (see e.g.\cite{integrabl}), 
\begin{eqnarray}
& & \frac1{N!}\oint\prod_{i=1}^N\frac{dz_i}{2\pi i}\ \sum_{\sigma\in S_N}(-1)^{\sigma}
    \prod_{k=1}^N {\cal K}(z_k,z_{\sigma(k)}) \nonumber \\
&=& \frac1{N!}{\sum_{\{d_l\}}}'\prod_{l=1}^\infty {((-1)^{l+1})}^{d_l}
    \frac{N!}{\prod_{l=1}^\infty(l!)^{d_l}d_l!}\prod_{l=1}^\infty\left( \frac{l!}{l} \right)^{d_l}
    \prod_{l=1}^\infty \left( \mbox{Tr}{\cal K}^l \right)^{d_l} \nonumber \\
&=& {\sum_{\{d_l\}}}'\prod_{l=1}^\infty
    \frac1{d_l!}\left( \frac{(-1)^{l+1}}{l}\mbox{Tr}{\cal K}^l \right)^{d_l}, 
\end{eqnarray}
where $\sum'$ represents the sum of $d_l$ with the constraint $\sum_{l=1}^\infty ld_l=N$. 
Then the grand canonical partition function is 
\begin{eqnarray}
Z(\lambda,\tilde{\lambda};\mu) 
&=& \sum_{N=0}^\infty e^{2\pi R\mu N}Z_N(\lambda,\tilde{\lambda}) \nonumber \\
&=& \sum_{\{d_l\}}\prod_{l=1}^\infty 
    \frac1{d_l!}\left( \frac{(-1)^{l+1}}{l}e^{2\pi R\mu l}\mbox{Tr}{\cal K}^l \right)^{d_l} 
    \nonumber \\
&=& \prod_{l=1}^\infty \exp\left[ \frac{(-1)^{l+1}}{l}e^{2\pi R\mu l}\mbox{Tr}{\cal K}^l \right] 
    \nonumber \\
&=& \exp\left[ \mbox{Tr}\log(1+e^{2\pi R\mu}{\cal K}) \right] \nonumber \\
&=& \mbox{Det}(1+e^{2\pi R\mu}{\cal K}). 
\end{eqnarray}

\vspace{5mm}
\section{Free field representation} \label{CFT}

The partition function (\ref{Z_N}) can be rewritten in terms of the free boson 
CFT\cite{Yin}. 
Define two free bosons 
\begin{equation}
\varphi_\alpha(z) = \hat{q}_\alpha+\hat{p}_\alpha\log z+\sum_{m\ne0}\frac{H_{\alpha,n}}{n}z^{-n}, 
\end{equation}
where $\alpha=1,2$ and 
\begin{equation}
[\hat{p}_\alpha,\hat{q}_\beta] = \delta_{\alpha\beta}, \hspace{5mm} 
[H_{\alpha,m},H_{\beta,n}] = m\delta_{\alpha\beta}\delta_{m+n,0}. 
\end{equation}
Two linear combinations $\phi(z), \tilde{\phi}(z)$ of $\varphi_\alpha(z)$, where 
\begin{eqnarray}
\phi(z) &=& \varphi_1(zq^{1/2})-\varphi_2(zq^{-1/2}), \\
\tilde{\phi}(z) &=& \varphi_1(zq^{-1/2})-\varphi_2(zq^{1/2}), 
\end{eqnarray}
are relevant for rewriting (\ref{Z_N}), and one can show that 
\begin{eqnarray}
& & \langle l|e^{-\sum_{n=1}^\infty(t_nH_n+\tilde{t}_n\tilde{H}_n)}
    \left(\oint\frac{dz}{2\pi i}\oint\frac{d\tilde{z}}{2\pi i}
    :e^{\phi(z)}::e^{-\tilde{\phi}(\tilde{z})}:\right)^N
    e^{\sum_{n=1}^\infty(t_{-n}H_{-n}+\tilde{t}_{-n}\tilde{H}_{-n})}|l\rangle 
      \nonumber \\
&=& (-1)^N(N!)^2q^{N(l_1+l_2)}f_l(t,\tilde{t})
    Z_N(\lambda,\tilde{\lambda}) 
\end{eqnarray}
where $l={l_1, l_2}$ collectively denotes the eigenvalues of $\hat{p}_\alpha$,
$f=l_1-l_2$ and 
\begin{equation}
f_l(t,\tilde{t}) = \langle l|e^{-\sum_{n=1}^\infty(t_nH_n+\tilde{t}_n\tilde{H}_n)}
                 e^{\sum_{n=1}^\infty(t_{-n}H_{-n}+\tilde{t}_{-n}\tilde{H}_{-n})}|l\rangle. 
\end{equation}
We have defined that 
\begin{equation}
H_{\alpha,n}|l\rangle = 0 \hspace{5mm} (n>0), \hspace{5mm} 
\hat{p}_\alpha|l\rangle = l_\alpha|l\rangle, 
\end{equation}
and $H_n,\tilde{H}_n$ are defined by the following expression of $\phi(z)$, 
\begin{equation}
\phi(z) = \hat{q}+\hat{p}\log z+\sum_{m\ne0}\frac{H_{n}}{n}z^{-n}, 
\end{equation}
and similar for tilder. 

Let us define the following quantity 
\begin{equation}
T_l(t,\tilde{t}) 
= f_l(t,\tilde{t})\sum_{N=0}^\infty e^{2\pi R\mu N}q^{N(l_1+l_2)}Z_N(\lambda,\tilde{\lambda}). 
\end{equation}
By the momentum conservation, this can be written as 
\begin{equation}
T_l(t,\tilde{t}) 
= \langle l|e^{-\sum_{n=1}^\infty(t_nH_n+\tilde{t}_n\tilde{H}_n)}g(\mu)
  e^{\sum_{n=1}^\infty(t_{-n}H_{-n}+\tilde{t}_{-n}\tilde{H}_{-n})}|l\rangle, 
\end{equation}
where 
\begin{equation}
g(\mu) = \exp\left( e^{\pi R\mu}\oint\frac{dz}{2\pi i}(e^{\phi(z)}-e^{-\tilde{\phi}(z)}) \right). 
\end{equation}
There are two tensor operators 
\begin{equation}
\sum_{\alpha,\beta}m_{\alpha\beta}\oint \frac{dz}{2\pi i}
e^{\varphi_\alpha(zq^{1/2})}\otimes e^{-\varphi_\beta(zq^{-1/2})}, \hspace{5mm} 
m=\sigma^1, \sigma^3, 
\end{equation}
which commute with $g(\mu)\otimes g(\mu)$. 
This commuting property leads to the following identities 
\begin{eqnarray}
& & \sum_{\alpha,\beta}m_{\alpha\beta}\oint\frac{dz}{2\pi i}
\langle l|e^{-\sum_{n=1}^\infty(t_nH_n+\tilde{t}_n\tilde{H}_n)}e^{\varphi_\alpha(z)}g(\mu)
  e^{\sum_{n=1}^\infty(t_{-n}H_{-n}+\tilde{t}_{-n}\tilde{H}_{-n})}|k\rangle \nonumber \\
& & \times\langle l'|e^{-\sum_{n=1}^\infty(t'_nH_n+\tilde{t}'_n\tilde{H}_n)}e^{-\varphi_\beta(z)}g(\mu)
  e^{\sum_{n=1}^\infty(t'_{-n}H_{-n}+\tilde{t}'_{-n}\tilde{H}_{-n})}|k'\rangle
  \nonumber \\
&=& \sum_{\alpha,\beta}m_{\alpha\beta}\oint\frac{dz}{2\pi i}
\langle l|e^{-\sum_{n=1}^\infty(t_nH_n+\tilde{t}_n\tilde{H}_n)}g(\mu)e^{\varphi_\alpha(z)}
  e^{\sum_{n=1}^\infty(t_{-n}H_{-n}+\tilde{t}_{-n}\tilde{H}_{-n})}|k\rangle \nonumber \\
& & \times\langle l'|e^{-\sum_{n=1}^\infty(t'_nH_n+\tilde{t}'_n\tilde{H}_n)}g(\mu)e^{-\varphi_\beta(z)}
  e^{\sum_{n=1}^\infty(t'_{-n}H_{-n}+\tilde{t}'_{-n}\tilde{H}_{-n})}|k'\rangle. \nonumber \\
\end{eqnarray}
These equations provide infinite number of differential equations which contain quantities 
\begin{equation}
T_{lk}(t,\tilde{t}) 
= \langle l|e^{-\sum_{n=1}^\infty(t_nH_n+\tilde{t}_n\tilde{H}_n)}g(\mu)
  e^{\sum_{n=1}^\infty(t_{-n}H_{-n}+\tilde{t}_{-n}\tilde{H}_{-n})}|k\rangle, 
\end{equation}
and their derivatives. 
Since the operator $g(\mu)$ commutes with $\hat{p}_1+\hat{p}_2$, $T_{lk}(t,\tilde{t})$ vanish unless  
$l_1+l_2 = k_1+k_2$, but since 
it does not commute with $\hat{p}_1-\hat{p}_2$, $T_{lk}(t,\tilde{t})$ is 
otherwise non-vanishing in general, 
and the system of equations is more complicated than that obtained from 
the Hirota bilinear equation for an ordinary integrable system. 

One way to simplify the situation is to represent $T_{lk}(t,\tilde{t})$ in terms of 
$T_{l}(t,\tilde{t})$. 
Suppose that $l_1=k_1+m, l_2=k_2-m$ for a positive integer $m$. 
Then one can show that 
\begin{equation}
T_{lk}(t,\tilde{t}) = {\cal I}_m\circ{\cal I}_{m-1}\circ\cdots\circ{\cal I}_1\left( 
                      T_k^{(m)}(t,\tilde{t}) \right), 
\end{equation}
where 
\begin{equation}
T_{k}^{(m)}(t,\tilde{t}) 
= \langle l|e^{-\sum_{n=1}^\infty(t_nH_n+\tilde{t}_n\tilde{H}_n)}
  \left( \oint\frac{dz}{2\pi i}e^{\phi(z)} \right)^mg(\mu)
  e^{\sum_{n=1}^\infty(t_{-n}H_{-n}+\tilde{t}_{-n}\tilde{H}_{-n})}|k\rangle, 
\end{equation}
and ${\cal I}_m$ is an integral transformation defined by 
\begin{equation}
{\cal I}_m(f(\mu)) = 2\pi R\ e^{m\pi R\mu}\int_\mu^\infty d\mu'e^{-\pi R(m+1)\mu'}f(\mu'). 
\end{equation}
$T_{k}^{(m)}(t,\tilde{t})$ can be written in terms of $T_{k}(t,\tilde{t})$ by moving all $e^{\phi(z)}$ 
to the left, and therefore we finally obtain a set of infinite number of integral-differential 
equations for $T_{k}(t,\tilde{t})$. 

\vspace{1cm}

\section{Proof of the $\lambda$-independence of the bosonic solution} \label{proof}

\vspace{5mm}

The second derivative of the free energy of the c=1 bosonic string at the self-dual radius is 
\begin{equation}
\chi_b(\mu) = -\log\mu+\sum_{N=1}^\infty\frac{2N-1}{2N}|B_{2N}|\mu^{-2N}. 
\end{equation}
To see that this is really a solution of the Toda equation (\ref{diff}), it is sufficient to see 
\begin{equation}
\left(\frac{\sin(\partial_\mu/2)}{\partial_\mu/2}\right)^2\chi_b(\mu)=-\log\mu. 
   \label{eq}
\end{equation}
It is more convenient to consider 
\begin{equation}
\left(\frac{\sin(\partial_\mu/2)}{\partial_\mu/2}\right)^2\partial_\mu\chi_b(\mu)=-\frac1\mu,  
   \label{eq2}
\end{equation}
and (\ref{eq}) follows from this. 

One can show that 
\begin{eqnarray}
& & \left(\frac{\sin(\partial_\mu/2)}{\partial_\mu/2}\right)^2\partial_\mu\chi_b(\mu) \nonumber \\
&=& -2\sum_{K=0}^\infty(-1)^K(2K)!\mu^{-2K-1}\sum_{k=0}^K\frac{|(2k-1)|B_{2k}}{(2k)!(2K-2k+2)!}. 
\end{eqnarray}
To evaluate the sum, let us calculate a related sum 
\begin{equation}
f(x)=\sum_{K=0}^\infty\sum_{k=0}^K\frac{(2k-1)B_{2k}}{(2k)!(2K-2k+2)!}x^{2K}, 
\end{equation}
which is 
\begin{eqnarray}
f(x) 
&=& \sum_{k=0}^\infty\frac{(2k-1)B_{2k}}{(2k)!}x^{2k}\sum_{l=0}^\infty\frac{x^{2l}}{(2l+2)!} 
    \nonumber \\
&=& -\frac{x^2e^x}{(e^x-1)^2}\frac{\cosh x-1}{x^2} \nonumber \\
&=& -\frac12. 
\end{eqnarray}
(\ref{eq2}) follows from this result.

\vspace{1cm}

\section*{Acknowledgments}
 We thank Piljin Yi for the participation of our project at the initial 
stage and intensive discussions throughout the work. The work of J.P. was supported by
Korea Science and Engineering Foundation (KOSEF) Grant R01-2004-000-10526-0 and by
POSTECH BSRI research fund 1RB0410601.

\end{document}